\documentclass[aps,prl,reprint,
amsmath,amssymb,
groupedaddress
showpacs]{revtex4-1}

\usepackage{graphicx}
\usepackage{dcolumn}
\usepackage{bm}
\usepackage{color}
\usepackage{ulem}

\begin{document}

\title{
Hidden Multiple-spin Interactions 
as an Origin of Spin Scalar Chiral Order 
in Frustrated Kondo Lattice Models
}

\author{Yutaka Akagi$^1$\thanks{E-mail address: akagi@aion.t.u-tokyo.ac.jp}, Masafumi Udagawa$^{1,2}$, and Yukitoshi Motome$^1$
}

\affiliation{$^1$Department of Applied Physics, University of Tokyo, Tokyo 113-8656, Japan\\
$^2$Max-Planck-Institut f\"{u}r Physik komplexer Systeme, 01187 Dresden, Germany}

\date{\today}

\begin{abstract}
We reveal the significance of kinetic-driven multiple-spin interactions 
hidden in geometrically-frustrated Kondo lattice models. 
Carefully examining the perturbation in terms of the spin-charge coupling up to the fourth order, 
we find that a positive biquadratic interaction is critically enhanced  
and plays a crucial role on stabilizing a spin scalar chiral order 
near 1/4 filling in a triangular lattice case. 
This is a generalized Kohn anomaly, 
appearing only when the second-order perturbation is inefficient 
because of the degeneracy under frustration. 
The mechanism is potentially common to frustrated spin-charge coupled systems,  
leading to emergence of unusual magnetic orders.
\end{abstract}


\maketitle

Localized spins in metal interact with each other 
via effective interactions mediated by the kinetic motion of itinerant electrons. 
An underlying mechanism is theoretically established 
as the Ruderman-Kittel-Kasuya-Yosida (RKKY) interaction~\cite{Ruderman1954,Kasuya1956,Yosida1957}. 
The precise form of the interaction sensitively depends on the electronic structure of the system 
as well as the coupling between charge and spin. 
This offers the possibility of various magnetic orderings in the spin-charge coupled systems.
 
Recently, unusual magnetic orderings have been of intensive research interest 
in geometrically-frustrated spin-charge coupled systems. 
In particular, a spin scalar chiral ordering has attracted considerable attention, 
since it underlies numerous fascinating phenomena, 
such as the unconventional anomalous Hall effect~\cite{Ohgushi2000,Taguchi2001}.
As a prominent example, 
a scalar chiral ordering with a four-sublattice noncoplanar spin configuration 
was discussed for the Kondo lattice model on a  
triangular lattice~\cite{Martin2008,Akagi_2010,Kumar_Brink_2010,Kato_2010}.
The particular order appears in two regions, near 3/4 filling and 1/4 filling~\cite{Akagi_2010}.  
While the former was predicted on the basis of 
the perfect nesting of the Fermi surface~\cite{Martin2008}, 
the latter is unexpected from 
the nesting scenario. 
Surprisingly, the 1/4 filling phase is stable up to much larger spin-charge coupling 
than the 3/4 filling one~\cite{Akagi_2010}. 
The robustness of the noncoplanar order near 1/4 filling casts a fundamental question on 
the mechanism of chiral ordering.

In this Letter, 
we point out the significance of kinetic-driven multiple-spin interactions for the noncoplanar ordering
under geometrical frustration. 
Carefully examining 
the perturbation in terms of the spin-charge coupling 
up to the fourth order, we find that  
the second-order RKKY contributions are degenerate 
among several different magnetic orderings because of the frustration, 
whereas the degeneracy is lifted by  
fourth-order contributions.
Among several terms, a positive biquadratic interaction is critically enhanced and plays a crucial role in 
the emergence of the noncoplanar order.  
We ascribe the mechanism of the critical enhancement to a 
Fermi surface effect --- a generalized Kohn anomaly at a higher level of perturbation. 
It opens a local energy gap at several points on the Fermi surface, 
which develops continuously 
to a full gap as increasing the spin-charge coupling 
and approaching 1/4 filling. 
This naturally explains the robustness of the chiral ordered phase.

\begin{figure}[b]
\begin{center}
\includegraphics[width=8.5cm]{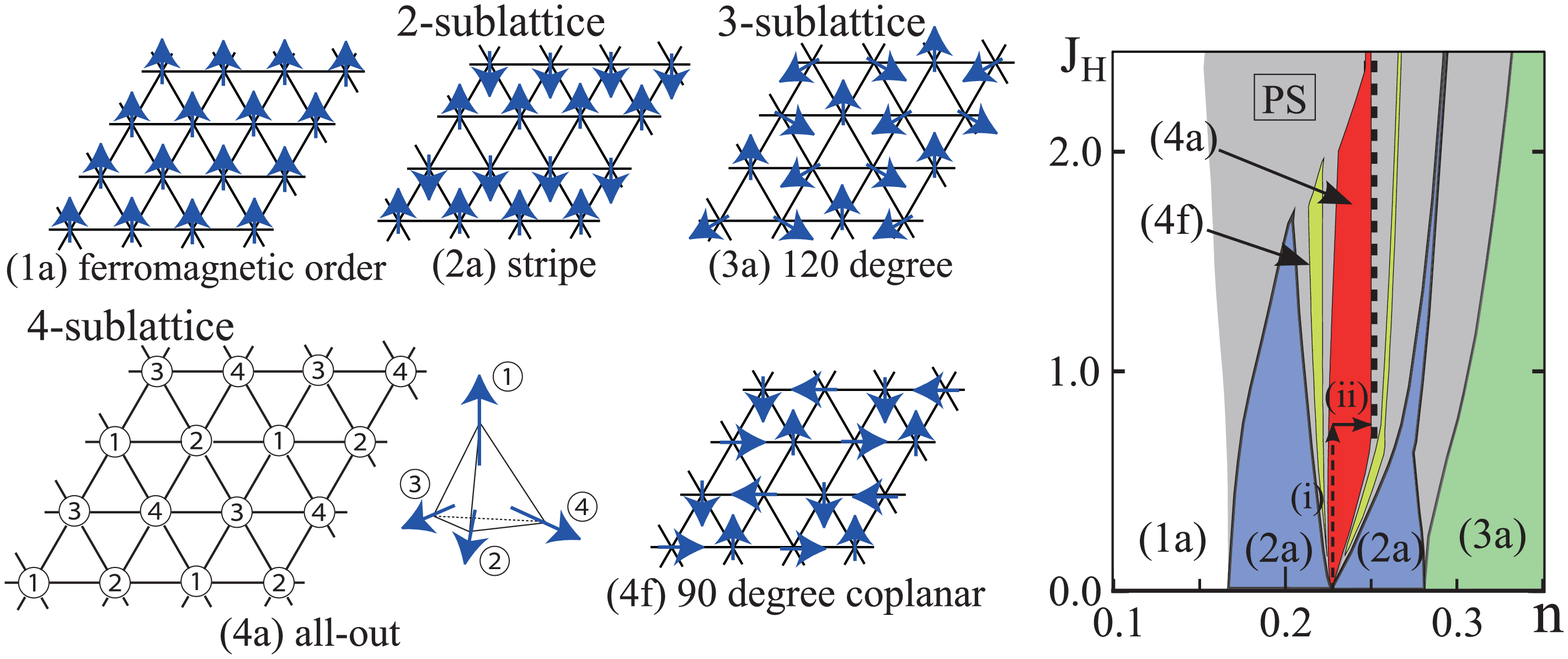}
\end{center}
\caption{(Color online)
Ordering patterns of localized spins considered in the perturbation calculations:
(1a) ferromagnetic, 
(2a) two-sublattice collinear stripe, 
(3a) three-sublattice $120^{\circ }$ noncollinear,  
(4a) four-sublattice all-out,   
and (4f) four-sublattice coplanar $90^{\circ }$ flux orders. 
The right panel is the 
phase diagram of Hamiltonian (\ref{eq:H}) near 1/4 filling. 
Arrows (i) and (ii) indicate the parameter changes used in Fig.~\ref{origin-of-biquadratic-interaction}(d). 
PS represents a phase-separated region and 
the bold dotted line at 1/4 filling denotes an insulating (4a) state. 
See also Figs.~1 and 3 in Ref.~\cite{Akagi_2010}.
}
\label{order-pattern}
\end{figure}

We consider the Kondo lattice model on a triangular lattice, 
following the previous studies~\cite{Martin2008,Akagi_2010,Kato_2010,AkagiPreprint}. 
The essential result is common to other frustrated lattices, as mentioned later. 
The Hamiltonian is given by 
\begin{equation}
{\cal H}=-t \!\! \sum_{\langle i,j \rangle,\alpha } \!\! ( c^{\dagger}_{i,\alpha } c_{j,\alpha }+\mathrm{h.c.}) 
-J_{\rm H} \! \sum_{i,\alpha ,\beta } \!
    c^{\dagger}_{i,\alpha } \boldsymbol{\sigma}_{\alpha \beta }  c_{i,\beta } \cdot \mathbf{S}_i,  
\label{eq:H}
\end{equation}
where $c^{\dagger}_{i,\alpha}$ ($c_{i,\alpha}$) is a creation (annihilation) operator of conduction electron at
site $i$ with spin $\alpha$, $\boldsymbol{\sigma}$ is the Pauli matrix, 
and ${\mathbf S}_i$ is a localized moment. 
The sum in the first term is taken over the nearest-neighbor sites on the triangular lattice. 
The sign of the spin-charge coupling $J_{\rm H}$ does not matter, 
since we assume ${\mathbf S}_i$ to be a classical spin. 

Given a specific spin configuration $\{ \mathbf{S}_i \}$, it is possible to obtain the exact free energy $F 
=-T\sum_{a}\log(1 + \exp(-\beta E_a))$ 
from the eigenvalues of $\mathcal{H}$, $\{E_a\}$ 
($T$ is temperature and $\beta=1/T$). 
In the previous study, 
from the comparison of 
the ground state energies, 
two of the authors discovered a scalar chiral phase 
with a four-sublattice noncoplanar spin order 
[Fig.~\ref{order-pattern}(4a)] 
near 1/4 filling~\cite{Akagi_2010}. 
As shown in the right panel of Fig.~\ref{order-pattern}, 
the chiral order sets in by an 
infinitesimal $J_{\rm H}$ from slightly less than 1/4 filling, 
and extended to 
a large $J_{\rm H}$ insulating region at 1/4 filling~\cite{Akagi_2010}.

In order to elucidate the origin of the robust noncoplanar ordering, 
we consider the perturbation in terms of $J_{\rm H}$. 
The analysis provides coherent understanding 
of the phase diagram not only in the small $J_{\rm H}$ region but in the 
large $J_{\rm H}$ insulating region, as we will 
discuss later. 
Taking the second term in eq.~(\ref{eq:H}) 
as the perturbation (${\cal H}'$) to the first term (${\cal H}_0$),  
we use the standard perturbation theory to 
calculate the energy for various ordering patterns.  
We here consider a collection of ordered states up to four-sublattice ordering, 
which appeared in the phase diagram in the small $J_{\rm H}$ region; 
namely, (1a) ferromagnetic, 
(2a) two-sublattice stripe,  
(3a) three-sublattice 120$^\circ$ coplanar,  
(4a) four-sublattice all-out,  
and (4f) four-sublattice flux orders, 
as shown in Fig.~\ref{order-pattern}.  
The free energy is expanded in the form 
\begin{align} 
F& -F_0=-T\log {\Big\langle {\cal T} 
\exp{\Big(-\int_{0}^{\beta } {\cal H}'(\tau ) d\tau  \Big)} \Big\rangle_{{\rm con}}} \notag\\
&=-\frac{T}{2!}\int_{0}^{\beta } d\tau _1 \int_{0}^{\beta } d\tau _2 \big\langle {\cal T}  
{\cal H}'(\tau _1) {\cal H}'(\tau _2) \big\rangle_{{\rm con}}  \notag\\
 & -\frac{T}{4!}\int_{0}^{\beta } d\tau _1 \cdots \int_{0}^{\beta } d\tau _4 \big\langle {\cal T} 
 {\cal H}'(\tau _1) \cdots {\cal H}'(\tau _4) \big\rangle_{{\rm con}} +\cdots\nonumber\\
 &= F^{(2)} + F^{(4)} + \cdots, \label{eq:perturbation}
\end{align}
where $F_0$ is the free energy obtained from  
${\cal H}_0$, $\tau$ is the imaginary time,
and ${\cal T}$ represents time-ordered product. 
$\langle \cdots \rangle_{\rm con}$ means the average 
over the connected diagrams.
From the spin rotational symmetry, 
odd order terms do not appear in the expansion. 

\begin{figure}[t]
 \begin{center}
  \includegraphics[width=8cm]{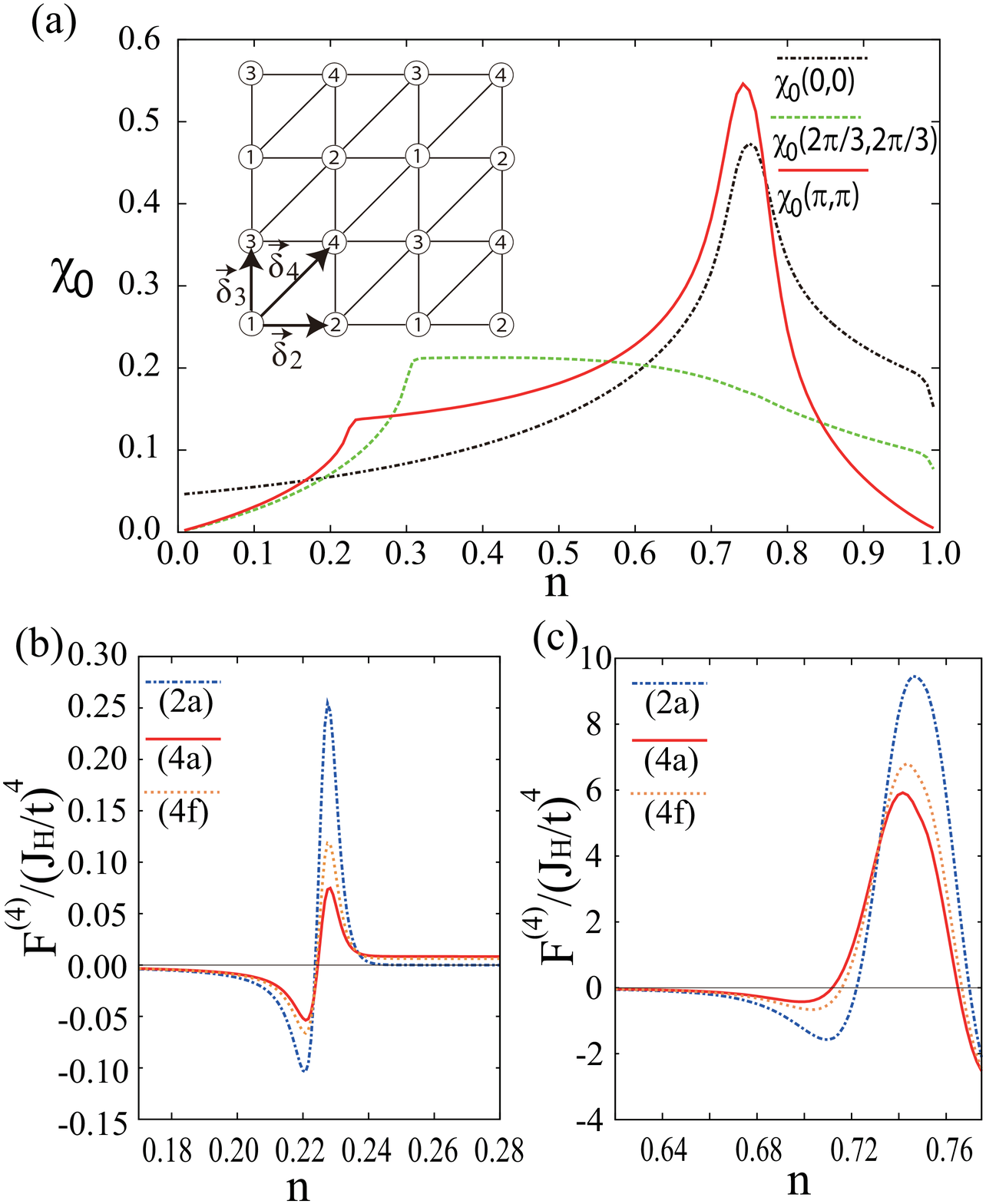}
 \end{center}
 \caption{(Color online) 
(a) $\chi _0(\mathbf{Q})$ in eq.~(\ref{eq:second_compare})
 as a function of $n$. 
 The inset represents a square lattice with diagonal bonds which is topologically equivalent to the triangular lattice.
(b), (c) The fourth-order free energies
of (2a), (4a), and (4f) states in the region near (b) $n\sim 1/4$ and (c) $n\sim 3/4$.
In the calculation, we take $800\times 800$ grid points in the 
extended first Brillouin zone and $T=0.03t$. 
}
 \label{Fig:perturbation}
\end{figure}

By decoupling the brackets $\langle \cdots \rangle_{\rm con}$ with the Wick's theorem, each term in  
eq.~(\ref{eq:perturbation}) can be written in a compact form.
The second-order term gives the well-known RKKY form~\cite{Ruderman1954,Kasuya1956,Yosida1957}; 
$
F^{(2)} 
= -J_{\rm H}^2 \sum_{\mathbf q} 
|\mathbf{S}_{\mathbf q}|^2\chi _0({\mathbf q})$,  
where the bare magnetic susceptibility is defined as 
$
\chi _0({\mathbf q}) =
-\frac{1}{N}\sum_{\mathbf{k}} 
\{
f_\textit{F}(\epsilon _{\mathbf{k}-\mathbf{q}})-f_\textit{F}(\epsilon _{\mathbf{k}})
\}/( 
\epsilon _{\mathbf{k}-\mathbf{q}}-\epsilon _{\mathbf{k}} 
)$, 
and $\mathbf{S}_{\mathbf q} = \frac{1}{N}\sum_{j} \mathbf{S}_j 
e^{i{\mathbf q} \cdot {\mathbf r}_j}$
($f_\textit{F}$ is the Fermi distribution function, $\epsilon _{\mathbf{k}}$ is an eigenvalue of ${\cal H}_0$ in the momentum basis, and $N$ is the number of sites).
Specifically, for each magnetic order, $F^{(2)}$ can be rewritten as
\begin{eqnarray}
F^{(2)} = -J_{\rm H}^2\times\left\{\begin{array}{ll}
\chi_0(0, 0) & \text{for (1a)}\\
\chi_0(2\pi/3, 2\pi/3) & \text{for (3a)}\\
\chi_0(\pi, \pi) & \text{for (2a)(4a)(4f).}
\end{array}\right.
\label{eq:second_compare}
\end{eqnarray}
Here, we take the triangular lattice in the form of a square lattice, 
as shown in inset of Fig.~\ref{Fig:perturbation}(a). 
Note that the (2a), (4a), and (4f) states have the same energy within the second-order 
perturbation.

The fourth-order term in eq.~(\ref{eq:perturbation})  
is calculated in a similar manner. 
For example, the result for the four-sublattice orders is 
summarized into a form of effective multiple-spin interactions as 
\begin{align} 
&F^{(4)} 
=\frac{T}{2}\left( \frac{J_{\rm H}}{4}\right) ^4 \sum_{\omega _p} \sum_{\mathbf{k}} \sum_{j_1, j_2, j_3, j_4=1\sim 4} \sum_{\mathbf{Q}_{l}, \mathbf{Q}_{m}, \mathbf{Q}_{n}} \notag\\
 & e^{i{\boldsymbol{\delta}} _{j_1} \cdot \left( \mathbf{Q}_{l}+\mathbf{Q}_{m}+\mathbf{Q}_{n} \right)} e^{-i{\boldsymbol{\delta}} _{j_2} \cdot \mathbf{Q}_{l}} e^{-i{\boldsymbol{\delta}} _{j_3} \cdot \mathbf{Q}_{m}} e^{-i{\boldsymbol{\delta}} _{j_4} \cdot \mathbf{Q}_{n}}  \times \notag\\
 & G_{\mathbf{k}}^0(\omega _p) G_{\mathbf{k}+\mathbf{Q}_{n}}^0(\omega _p) 
G_{\mathbf{k}+\mathbf{Q}_{m}+\mathbf{Q}_{n}}^0(\omega _p) G_{\mathbf{k}+\mathbf{Q}_{l}+\mathbf{Q}_{m}+\mathbf{Q}_{n}}^0 (\omega _p) \notag\\ 
&\quad  \times \left[ ( \mathbf{S}_{j_1} \cdot \mathbf{S}_{j_2} ) ( \mathbf{S}_{j_4} \cdot \mathbf{S}_{j_3} )
+( \mathbf{S}_{j_1} \times \mathbf{S}_{j_2} ) \cdot ( \mathbf{S}_{j_4} \times \mathbf{S}_{j_3} ) \right] \notag\\
&= J_{\rm H}^4 \!\!\!
\sum_{1\leq i<j \leq 4}\bigl[ J\mathbf{S}_{i} \cdot \mathbf{S}_{j} +B(\mathbf{S}_{i} \cdot \mathbf{S}_{j})^2 +B^{\ast }(\mathbf{S}_{i} \times  \mathbf{S}_{j})^2 \bigr] \notag\\
& + \!\! \sum_{i\neq j\neq k}\bigl[ M( \mathbf{S}_{i} \cdot \mathbf{S}_{j})( \mathbf{S}_{i} \cdot \mathbf{S}_{k}) +M^{\ast }( \mathbf{S}_{i} \times  \mathbf{S}_{j}) \cdot ( \mathbf{S}_{i} \times  \mathbf{S}_{k}) \bigr] \notag\\
& + \!\! \sum_{i\neq j\neq k\neq l}\bigl[ L( \mathbf{S}_{i} \cdot \mathbf{S}_{j})( \mathbf{S}_{k} \cdot \mathbf{S}_{l}) +L^{\ast }( \mathbf{S}_{i} \times  \mathbf{S}_{j}) \cdot ( \mathbf{S}_{k} \times  \mathbf{S}_{l}) \bigr]. 
\label{effective-Hamiltonian}
\end{align}
Here the sum of ${\mathbf{Q}_{l,m,n}}$ runs over 
$(0,0)$, $(\pi,0)(=\mathbf{Q}_a)$, $(0,\pi)(=\mathbf{Q}_b)$, and $(\pi,\pi)(=\mathbf{Q}_c)$ 
($\mathbf{Q}_a$, $\mathbf{Q}_b$, and $\mathbf{Q}_c$ are 
four-sublattice ordering wave vectors~\cite{Martin2008}). 
The indices $i,j$ denote the four sublattices, and 
$\boldsymbol{\delta}_i = 
(0, 0)$, $(1, 0)$, $(0, 1)$, and $(1 ,1)$,
 respectively [see inset of Fig.~\ref{Fig:perturbation}(a)]. 
$G_{\mathbf{k}}^0(\omega_p)=\{i\omega _p-(\epsilon _{\mathbf{k}}-\mu)\}^{-1}$
 is the noninteracting Green's function, 
where $\omega _p$ is the Matsubara frequency and 
 $\mu$ is a chemical potential. 
Thus, the fourth-order contribution includes three-spin and four-spin effective interactions in addition to two-spin interactions.
Each coefficient is given by the sum of four Green's functions products.
   
First let us discuss the result of second-order perturbation. 
The second-order term is proportional to $\chi_0$, 
as shown in eq.~(\ref{eq:second_compare}). 
The comparison of $\chi_{0}$ as a function of the electron density $n$ is 
presented in Fig.~\ref{Fig:perturbation}(a).
 Since the magnetic order with largest $\chi_{0}$ is stabilized 
when $J_{\rm H}$ is turned on, 
the result indicates that the ferromagnetic order appears for 
 $0.0 \leq n \lesssim 0.15$ and $0.8 \lesssim n \leq 1.0$,
 and the three-sublattice 120$^{\circ }$ coplanar order is realized for $0.3 \lesssim n \lesssim 0.55$.
See also Fig.~3 in Ref.~\cite{Akagi_2010}.
Meanwhile, $\chi _0(\pi, \pi)$ is the largest 
in the remaining regions, 
in which we have to consider the next fourth-order terms to lift the degeneracy
between (2a), (4a), and (4f) states~\cite{2nd_comment}.

The fourth-order free energies in the remaining regions for (2a), (4a), and (4f) 
 are indicated in Figs.~\ref{Fig:perturbation}(b) and \ref{Fig:perturbation}(c). 
The results show that (2a) two-sublattice stripe phase 
has the lowest fourth-order contribution 
in a wide range of $n$, while (4a) four-sublattice all-out order 
becomes lowest in narrow windows at $n\simeq 0.225$ and $n\simeq  0.75$. 
The stability of (4a) state at $n \simeq 0.75$ is explained by  
the perfect nesting of the Fermi surface at 3/4 filling~\cite{Martin2008}. 
On the other hand, the reason for the stabilization at $n\simeq 0.225$ is not obvious.

  \begin{figure}[!htbp]
 \begin{center}
  \includegraphics[width=7cm]{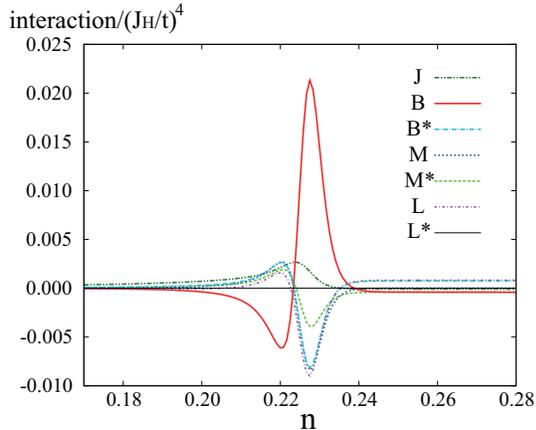}
 \end{center}
 \caption{(Color online) Coefficients of the different terms in 
 the fourth-order effective Hamiltonian [eq.~(\ref{effective-Hamiltonian})] 
as a function of the electron density $n$. 
}
 \label{triangular_4-th_effective-Hamiltonian_n0.25}
\end{figure}

In order to gain an insight into how the (4a) state is stabilized at $n \simeq 0.225$, 
we carefully analyze the fourth-order contributions in eq.~(\ref{effective-Hamiltonian}) term by term. 
We plot the coefficient of each term,  
the product of the Green's functions, 
as a function of $n$ in Fig.~\ref{triangular_4-th_effective-Hamiltonian_n0.25}.
The results show that all the coefficients show drastic changes around 
$n \simeq 0.225$: In particular, 
the coefficient of biquadratic interaction, $B$, shows a 
pronounced peak with the largest contribution. 
This indicates that the effective positive biquadratic interaction plays a dominant role  at $n \simeq 0.225$.
In general, the positive biquadratic interaction 
favors a noncollinear spin configuration. 
Particularly, among the degenerate (2a), (4a), and (4f) states, 
the (4a) noncoplanar all-out order is naturally selected by this interaction.

The emergence of the positive biquadratic interaction is remarkable, 
because the effective biquadratic interaction has a negative coefficient in most cases.
For example, in the Hubbard-type Hamiltonian, a fourth-order perturbation from the strong-coupling limit 
usually leads to a negative biquadratic interaction. 
Moreover, in localized spin systems, a spin-lattice coupling and 
 thermal/quantum fluctuations 
 also give rise to negative biquadratic interactions.
Our result indicates that an unusual positive biquadratic interaction 
is induced by the kinetic motion of electrons at a special filling.

 \begin{figure}[!htbp]
 \begin{center}
  \includegraphics[width=8.5cm]{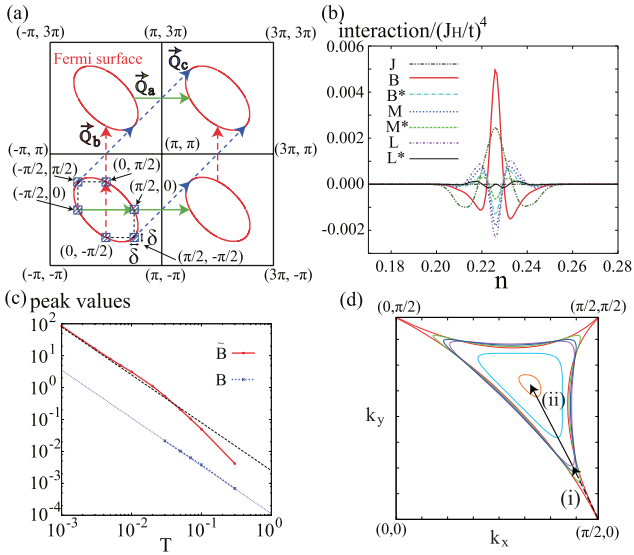}
 \end{center}
 \caption{(Color online) (a) The Fermi surface at $n=0.225$ in the extended Brillouin zone scheme.  
The connections of the Fermi surfaces by the four-sublattice wave vectors $\mathbf{Q}_{a}=(\pi, 0)$, $\mathbf{Q}_{b}=(0, \pi)$, $\mathbf{Q}_{c}=(\pi, \pi)$ are shown.
Six hatched squares 
in the first Brillouin zone indicate the patches 
used in the calculations of the data in (b). 
 The patches are $\delta \times \delta$ each, and centered 
at the special $\mathbf{k}$ points, ($\pm\pi/2,0$), ($0,\pm\pi/2$), and ($\pm\pi/2,\mp\pi/2$), which are connected with each other by the four-sublattice wave vectors.
(b) Coefficients of the effective four-spin interactions in eq.~(\ref{effective-Hamiltonian}) obtained by taking the $\mathbf{k}$ summation only within the patches 
in (a) with $\delta =\pi/40$. 
 (c) The peak values of 
 $\tilde{B}=T \sum_{\mathbf{k}} \sum_{\omega_p} 
[G_\mathbf{k}^0(\omega_p)]^2 
[G_{\mathbf{k} + \mathbf{Q}}^0 (\omega_p)]^2$ with patch $\delta =\pi/8$
as a function of $T$. 
The peak value of $B$ calculated by the integration over the 
entire first Brillouin zone is also plotted. 
Two straight lines are the fitting to $T^{-1.5}$.
(d) The Fermi surface evolution 
in the (4a) phase along the 
arrows in the right panel of Fig.~\ref{order-pattern}; 
(i) from $J_{\rm H}=0$ to $0.75$ (every $0.25$) at a fixed $n=0.2255$ 
and (ii) from $n=0.231$ to $0.249$ (every $0.006$) at a fixed $J_{\rm H}=0.75$. 
Only the first quadrant of the folded 
Brillouin zone is shown.
}
 \label{origin-of-biquadratic-interaction}
\end{figure}

Why is the positive biquadratic interaction enhanced here?
In order to elucidate the origin,  
let us look at the electronic state at $n\simeq 0.225$ in detail.
Figure~\ref{origin-of-biquadratic-interaction}(a) shows the Fermi surface at $n=0.225$, which is circular, with no substantial nesting tendency.
At first sight, this featureless Fermi surface does not lead to any type of singularity.
It, however, conceals an unexpected 
instability toward 
four-sublattice ordering.
 The key observation is that the ordering vectors of four sublattice order, $\mathbf{Q}_{a}=(\pi, 0)$, $\mathbf{Q}_{b}=(0, \pi)$, and $\mathbf{Q}_{c}=(\pi, \pi)$, 
connect the particular points on the Fermi surface 
when $n$ reaches $\simeq 0.225$; 
the tangents of the Fermi surface are parallel between the connected pairs --- see 
Fig.~\ref{origin-of-biquadratic-interaction}(a). 
These connections give rise to a singularity analogous to the $2k_F$-singularity or Kohn anomaly, observed in an isotropic electron gas~\cite{Kohn_1959}. 
It is worthy to note that, 
in contrast to the Kohn anomaly usually discussed in the second order perturbation~\cite{Kohn_1959}, 
the singularity considered here arises in the fourth-order perturbation~\cite{Lindhard_comment}.

The scenario is confirmed by numerically calculating the contributions only 
 from the parts of Fermi surface connected by $\mathbf{Q}_a$, $\mathbf{Q}_b$, and $\mathbf{Q}_c$.
We calculate the coefficients of the effective four-spin interactions in eq.~(\ref{effective-Hamiltonian}) by limiting the $\mathbf{k}$ summation 
 within the small hatched patches shown in Fig.~\ref{origin-of-biquadratic-interaction}(a). 
As shown in Fig.~\ref{origin-of-biquadratic-interaction}(b),  
the contributions from the small patches with $\delta =\pi/40$ lead to 
a positive and largest biquadratic interaction $B$, which well accounts for the result in Fig.~\ref{triangular_4-th_effective-Hamiltonian_n0.25}. 
This indicates that the anomalous behaviors in Fig.~\ref{triangular_4-th_effective-Hamiltonian_n0.25} 
 can be attributed to the particle-hole process within such small area in the momentum space.

The critical enhancement is, in fact, a divergence 
in the limit of $T \to 0$ in the perturbative calculations. 
The $T$ dependence is analytically calculated by approximating 
the connected fragments of the Fermi surfaces by simple biquadratic functions 
centered at the connected $\mathbf{k}$ points.  
The dominant contribution to the divergence comes from the terms proportional to $T \sum_{\mathbf{k}} \sum_{\omega_p} 
[G_\mathbf{k}^0(\omega_p)]^2[G_{\mathbf{k} + \mathbf{Q}}^0 (\omega_p)]^2$ ($\equiv \tilde{B}$) in eq.~(\ref{effective-Hamiltonian}); 
it leads to 
quick divergence $\propto$ $T^{-1.5}$, reflecting the fact that the tangents of the Fermi surface are parallel to each other at the connected points.
The details of the calculaions will be reported elsewhere. 
Indeed, the numerically calculated values of 
$\tilde{B}$ as well as $B$ 
increase with $\propto T^{-1.5}$ as $T \to 0$, 
as shown in Fig.~\ref{origin-of-biquadratic-interaction}(c). 

The divergence leads to a local gap formation at the connected points on the Fermi surface. 
Figure~\ref{origin-of-biquadratic-interaction}(d) shows the Fermi surface evolution 
in the (4a) phase, as we change the parameters ($n, J_{\rm H}$) along the arrows 
(i) and (ii) on the right panel of Fig.~\ref{order-pattern}. 
As soon as $J_{\rm H}$ is switched on, 
the Fermi surface is modified at the connected points 
reflecting the local gap opening. 
The gap extends to shrink the Fermi surface as
$J_{\rm H}$ increases and $n$ becomes closer to 1/4 filling; 
eventually, the Fermi surface disappears in the insulating state at 1/4 filling. 
This continuous change clearly demonstrates that 
the generalized Kohn anomaly revealed by the perturbative argument 
provides an essential ingredient for the robust chiral ordering including the insulating region.

Let us emphasize the importance of the geometrical frustration in this generalized Kohn anomaly.
Frustration leaves degeneracy at the level of second-order perturbation, 
and prepares the stage for the fourth-order term to select the type of magnetic order.  
In this sense, geometrical frustration is an indispensable prerequisite for the 
positive biquadratic interaction to play an active role.
In fact, when the frustration is reduced by introducing a spatial 
anisotropy in the triangular lattice structure, 
noncoplanar order is not stabilized within the four-sublattice order, 
even though the positive biquadratic interaction is present in the fourth-order terms:
the effect of positive biqadratic interaction is completely masked by the second-order RKKY interaction.
Meanwhile, for other frustrated lattices,
such as checkerboard, face-centered-cubic, and pyrochlore lattices, 
we found that similar four-sublattice order is commonly stabilized near 1/4 filling 
by the positive biqiuadratic interaction~\cite{Chern_2010}. 
This suggests that the present mechanism is robust widely in frustrated spin-charge coupled systems.

To summarize, we have carried out the perturbation expansion up to the fourth order in $J_{\rm H}/t$ and constructed 
the effective multiple-spin interactions for the ferromagnetic
 Kondo lattice model on a triangular lattice. Among the kinetic-driven 
four-spin interactions, we have found that a biquadratic interaction takes a large positive coefficient 
and increases divergently as $T \to 0$ near 1/4 filling. 
 This signals an instability toward a noncoplanar chiral phase near the 1/4 filling with a local gap formation.
The origin of large positive biquadratic interaction 
is ascribed to the Fermi surface connection by the ordering wave vectors of four sublattice
order, which we call the generalized Kohn anomaly. 
Our result provides comprehensive understanding of 
not only the emergence but also the robustness of the chiral ordering. 
The mechanism is universal in frustrated spin-charge coupled systems, 
possibly even when electron correlation between conduction electrons is included 
unless the Fermi surface is destroyed by electron correlation.

\acknowledgements
{
We acknowledge helpful discussions with Takahiro Misawa and Youhei Yamaji. 
This work was supported by Grants-in-Aid for Scientific Research 
(No. 19052008, 21340090, and 21740242), 
Global COE Program ``the Physical Sciences Frontier", 
the Strategic Programs for Innovative
Research (SPIRE), MEXT, and the Computational Materials Science
Initiative (CMSI), Japan.
}


\begin{thebibliography}{20}
\bibitem{2nd_comment}
At a general filling, we need to consider a larger magnetic unit cell, since $\chi _0(\mathbf{q})$ shows a 
peak at an incommensurate wave vector. 
At $n\simeq 0.225$, however, $\chi _0(\mathbf{q})$ has maxima at the four-sublattice ordering wave vectors, $\mathbf{q}=\mathbf{Q}_{a}$, $\mathbf{Q}_{b}$, and $\mathbf{Q}_{c}$. Therefore, our argument on the basis of four-sublattice unit cell
is plausible.
\bibitem{Lindhard_comment} 
We note that a singularity is already seen 
in the second-order perturbation, 
as expected from the usual Kohn anomaly: 
As shown in Fig.~\ref{Fig:perturbation}(a), a shoulder-like singularity of 
$\chi _0(\pi, \pi)$ at $n\simeq0.225$, 
analogous to the polarization function of a 
two-dimensional electron gas --- the so-called Lindhard function.
\end{thebibliography}
\end{document}